\documentclass{article}

\usepackage[preprint]{neurips_2023}
\setcitestyle{round} 

\usepackage[utf8]{inputenc} % allow utf-8 input
\usepackage[T1]{fontenc}    % use 8-bit T1 fonts
\usepackage{hyperref}       % hyperlinks
\usepackage{url}            % simple URL typesetting
\usepackage{booktabs}       % professional-quality tables
\usepackage{amsfonts}       % blackboard math symbols
\usepackage{nicefrac}       % compact symbols for 1/2, etc.
\usepackage{microtype}      % microtypography
\usepackage{xcolor}         % colors
\usepackage{pgfplots}       % Line plot
\usepackage{multirow}
\usepackage{adjustbox}
\usepackage{xurl}           % Line breaks in URLs
\usepackage{tabularx}

\newcolumntype{L}[1]{>{\raggedright\arraybackslash}p{#1}}
\newcolumntype{C}[1]{>{\centering\arraybackslash}p{#1}}
\newcolumntype{R}[1]{>{\raggedleft\arraybackslash}p{#1}}

\title{Comparative Analysis of Retrieval Systems \\ in the Real World}

\author{%
  \textbf{Dmytro Mozolevskyi} \And Waseem AlShikh \AND
    Writer, Inc. \\
  \texttt{\{dmytro,waseem\}@writer.com} \\
}
\begin{document}

\maketitle

% \begin{abstract}

% To do?
% \end{abstract}

\section{Introduction}

In the rapidly evolving landscape of information retrieval and natural language processing, the integration of advanced language models with search and retrieval systems has become a cornerstone for enhancing the quality and efficiency of data processing. This report presents a comprehensive analysis of various state-of-the-art methods that combine cutting-edge language models with sophisticated retrieval techniques. Our objective is to evaluate and compare these methods based on their performance in two critical aspects: accuracy, as measured by the RobustQA average score \citep{han-etal-2023-robustqa}, and efficiency, determined by the average response time.

We have explored a diverse range of methods including Azure Cognitive Search Retriever integrated with GPT-4,\footnote{\url{https://github.com/Azure-Samples/azure-search-openai-demo}} Pinecone's Canopy framework,\footnote{\url{https://www.pinecone.io/blog/canopy-rag-framework/}} various implementations of Langchain with Pinecone and different language models (OpenAI, Cohere),\footnote{\url{https://python.langchain.com/docs/integrations/vectorstores/pinecone}}\textsuperscript{,}\footnote{\url{https://python.langchain.com/docs/integrations/retrievers/cohere}} LlamaIndex with Weaviate Vector Store's hybrid search,\footnote{\url{https://docs.llamaindex.ai/en/stable/examples/vector_stores/WeaviateIndexDemo-Hybrid.html}} Google's RAG implementation on Cloud VertexAI-Search,\footnote{\url{https://python.langchain.com/docs/templates/rag-google-cloud-vertexai-search}} Amazon SageMaker's RAG,\footnote{\url{https://aws.amazon.com/blogs/machine-learning/question-answering-using-retrieval-augmented-generation-with-foundation-models-in-amazon-sagemaker-jumpstart/}} and a novel approach combining a graph search algorithm with a language model and retrieval awareness (Writer Retrieval).

The impetus for this analysis stems from the increasing demand for robust and responsive question-answering systems in various domains, from customer service to academic research. As the complexity of queries and the volume of information grows, it becomes imperative to not only retrieve relevant information quickly but also to ensure the precision and adaptability of the responses. The RobustQA metric \citep{han-etal-2023-robustqa}, a pivotal element in our analysis, offers a nuanced view of how well these systems perform under diverse paraphrasing of questions, reflecting real-world querying scenarios.

This report aims to provide insights into the strengths and weaknesses of each method, drawing comparisons that would help in understanding which combinations of technologies and approaches are best suited for specific applications. By delving into performance data, we aim to shed light on the current state of AI-driven search and retrieval systems and to pave the way for informed decisions in the deployment and development of these technologies.

\section{Background}

In the realm of natural language processing (NLP) and question-answering (QA) systems, the need for robust and reliable evaluation metrics has always been paramount. The RobustQA \citep{han-etal-2023-robustqa} metric is a significant step towards addressing this need. 

RobustQA is an innovative framework designed to evaluate the robustness of QA systems. It focuses on assessing how well these systems handle diverse paraphrasings of questions. This approach is crucial because real-world queries often come in various forms and styles, and a truly robust QA system must handle this variability effectively. 

The framework employs a novel methodology that goes beyond traditional evaluation metrics, which often rely on a limited set of question variations. RobustQA introduces a broader and more challenging set of paraphrased questions, thereby providing a more comprehensive and realistic assessment of a QA system's performance.

By incorporating RobustQA into our evaluation of different language model integrations with search and retrieval systems, we aim to gain deeper insights into their real-world effectiveness. This approach allows us to understand not only the accuracy of these systems in responding to queries but also their adaptability and resilience to varied linguistic expressions. Various statistics about the data in RobustQA are shown in \autoref{tab:robustqa-data}.

\begin{table}[t]
  \centering
\begin{adjustbox}{width=\textwidth}
    \begin{tabular}{c|llrrrl}
    \toprule
          & \textbf{Domain} & \textbf{Label} & \multicolumn{1}{l}{\textbf{\# Test Questions}} & \multicolumn{1}{l}{\textbf{\# Documents}} & \multicolumn{1}{l}{\textbf{\# Passages}} & \textbf{Data Source} \\
    \midrule
    \multicolumn{1}{l|}{\textbf{NQ}} & Wikipedia & [NQ]  & 3,610 & \multicolumn{1}{l}{-} & 21,015,324 & NQ \\
    \midrule
    \multirow{8}[2]{*}{\textbf{RobustQA}} & Web-search & [SE]  & 31,760 & 13,791,373 & 13,791,592 & SearchQA \\
          & Biomedical & [BI]  & 1,956  & 15,559,026 & 37,406,880 & BioASQ \\
          & Finance & [FI]  & 3,669  & 57,638 & 105,777 & FiQA \\
          & Lifestyle & [LI]  & 2,214  & 119,461 & 241,780 & LoTTE \\
          & Recreation & [RE]  & 2,096  & 166,975 & 315,203 & LoTTE \\
          & Technology & [TE]  & 2,115  & 638,509 & 1,252,402 & LoTTE \\
          & Science & [SC]  & 1,426  & 1,694,164 & 3,063,916 & LoTTE \\
          & Writing & [WR]  & 2,696  & 199,994 & 347,322 & LoTTE \\
    \bottomrule
    \end{tabular}%
\end{adjustbox}
  \caption{Table and caption recreated from \citep[Table 2]{han-etal-2023-robustqa}. Data summary: Natural Questions \citep[NQ]{kwiatkowski-2019} (top) v.s. RobustQA (bottom). \# Documents for NQ is missing because we directly use the passage split provided by \citep{karpukhin-etal-2020-dense}. Passages consists of 100 continuous tokens at most from the original documents.}
  \label{tab:robustqa-data}%
\end{table}%

\section{Experiments and Results}

In our empirical evaluation, we rigorously assessed the efficacy of eight different retrieval system configurations by scrutinizing two primary dimensions: the RobustQA metric score and the response time latency. Both metrics are salient indicators of the systems' proficiency in handling real-world query scenarios with swiftness and precision. The technical configurations tested ranged between different combinations of search architectures, integrated language models, and indexing techniques.

\textbf{RobustQA Score Estimation:}
The RobustQA score computation leverages a precision-oriented methodology that quantifies the systems' accuracy in retrieving the correct response against a corpus of paraphrased queries. This measure is calculated through an analytic comparison, where each system-generated response is benchmarked against a gold-standard dataset that encapsulates a broad spectrum of question paraphrases, reflecting realistic variance in linguistic expression. A high-resolution scoring mechanism, it accommodates shuffled question sequences to mitigate position bias and ensure comprehensiveness in evaluation.

\textbf{Response Time Measurement:}
This metric is quantified by recording the interval from the initiation of the query request to the point when a complete response set is rendered to the end user. Precision timers embedded within the system's operational pipeline log these intervals, capturing the raw computational throughput and data retrieval cadence. This temporal metric critically reflects the operational efficiency and scalability of the system, with adjustments made for network latency effects to ensure the metric purity.

Moving to the experimental setup, each retrieval system configuration was rigorously evaluated in a controlled test environment to ensure repeatability and consistency. The subsequent report meticulously tabulates these findings:

\begin{table}[t]
  \centering
    \begin{tabular}{L{8cm}R{2.2cm}R{2.2cm}}
    \toprule
    \textbf{Pipeline} & \textbf{RobustQA Avg. score} & \textbf{Avg. response time (secs)} \\
    \midrule
    Azure Cognitive Search Retriever + GPT4 + Ada & 72.36 & >1.0s \\
    Canopy (Pinecone) & 59.61 & >1.0s \\
    Langchain + Pinecone + OpenAI & 61.42 & <0.8s \\
    Langchain + Pinecone + Cohere & 69.02 & \textbf{<0.6s} \\
    LlamaIndex + Weaviate Vector Store - Hybrid Search & 75.89 & <1.0s \\
    RAG Google Cloud VertexAI-Search + Bison & 51.08 & >0.8s \\
    RAG Amazon SageMaker & 32.74 & <2.0s \\
    Graph search algorithm + LLM + Retrieval awareness (Writer Retrieval) & \textbf{86.31} & \textbf{<0.6s} \\
    \bottomrule
    \end{tabular}%
  \caption{Comparison of different pipelines on RobustQA in terms of both performance and response time. Best scores are in bold.}
  \label{tab:results}%
\end{table}%

\paragraph{Azure Cognitive Search Retriever + GPT4 + Ada\protect\footnote{\url{https://github.com/Azure-Samples/azure-search-openai-demo}}}

This combination shows a strong performance in terms of accuracy, indicating good integration between Azure's search capabilities and GPT-4's Ada model. The response time is over one second, suggesting a balance between thoroughness and speed.

\paragraph{Canopy Pinecone)\protect\footnote{\url{https://www.pinecone.io/blog/canopy-rag-framework/}}}

The lower score here could indicate limitations in either the Canopy framework's retrieval ability or its integration with Pinecone. The response time is similar to the Azure Cognitive Search setup.

\paragraph{Langchain + Pinecone + OpenAI\protect\footnote{\url{https://python.langchain.com/docs/integrations/vectorstores/pinecone}}}

This method shows a modest improvement in accuracy over Canopy, with a notably faster response time. The Langchain framework might be more efficient in querying or processing results.

\paragraph{Langchain + Pinecone + Cohere\protect\footnote{\url{https://python.langchain.com/docs/integrations/retrievers/cohere}}}

This setup shows a significant jump in accuracy and a further reduction in response time. The Cohere model's integration with Langchain and Pinecone seems particularly effective.

\paragraph{LlamaIndex + Weaviate Vecor Score - Hybrid Search\protect\footnote{\url{https://docs.llamaindex.ai/en/stable/examples/vector_stores/WeaviateIndexDemo-Hybrid.html}}}

This method is second best in terms of accuracy, suggesting that LlamaIndex's hybrid search approach effectively leverages Weaviate's vector store capabilities. The sub-one-second response time also indicates efficiency.

\paragraph{RAG Google Cloud VertexAI-Search + Bison\protect\footnote{\url{https://python.langchain.com/docs/templates/rag-google-cloud-vertexai-search}}}

This combination shows the lowest accuracy, which might be due to how RAG and Bison interact or how they leverage Google Cloud's VertexAI-Search. The response time is moderate.

\paragraph{RAG Amazon SageMaker\protect\footnote{\url{https://aws.amazon.com/blogs/machine-learning/question-answering-using-retrieval-augmented-generation-with-foundation-models-in-amazon-sagemaker-jumpstart/}}}

This setup has the lowest score and the longest response time, indicating possible inefficiencies or mismatches in the integration of RAG with Amazon SageMaker.

\paragraph{Graph search algorithm + LLM + Retrieval awareness (Writer Retrieval)}

This method outperforms all others in accuracy while maintaining a fast response time, highlighting the effectiveness of combining a graph search algorithm with a language model and retrieval awareness.

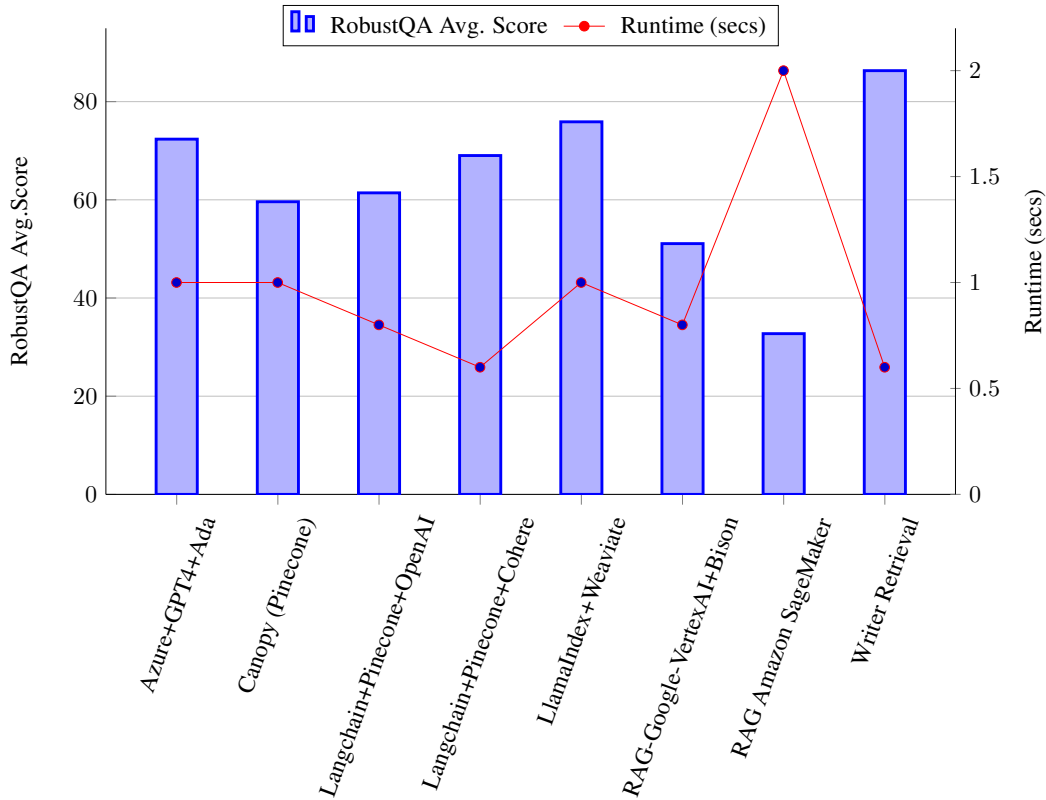
\begin{figure}
    \centering
    \begin{adjustbox}{width=\textwidth}
    \begin{tikzpicture}
    \pgfplotsset{lineplot/.style={blue,mark=*,sharp plot,line legend}}
        \begin{axis}[
            ybar,
            ymin = 0,
            axis x line* = bottom,
            axis y line* = left,
            ylabel= RobustQA Avg.Score,
            ymajorgrids = true,
            width= \textwidth,
            height = 0.6\textwidth,
            bar width = 0.6cm,
            xtick= {1,2,3,4,5,6,7,8},
            xticklabels= {Azure+GPT4+Ada, Canopy (Pinecone), Langchain+Pinecone+OpenAI, Langchain+Pinecone+Cohere, LlamaIndex+Weaviate, RAG-Google-VertexAI+Bison, RAG Amazon SageMaker, Writer Retrieval
            },
            x tick label style={rotate=70},
            legend columns=2,
            legend cell align=left,
            legend style={
                at={(0.5,1.05)},
                anchor=north,
                column sep=1ex}
            ]
            \addplot+[mark=none,very thick,label=barplot] coordinates {
                (1, 72.36)
                (2, 59.61)    
                (3, 61.42) 
                (4, 69.02) 
                (5, 75.89) 
                (6, 51.08) 
                (7, 32.74) 
                (8, 86.31) 
            };
                \addlegendentry{RobustQA Avg. Score};
                \addlegendimage{lineplot, red}
                \addlegendentry{Runtime (secs)}
        \end{axis} 
        \begin{axis}[
            axis y line=right,
            axis x line=none,
            width= \textwidth,
            height = 0.6\textwidth,
%            ymajorgrids = true,
            axis line style={-},
            ylabel = {Runtime (secs)},
            scaled y ticks = false,
            ymin=0, ymax=2.2,
        ]
         \addplot+[lineplot, red] coordinates {
                    (1, 1.0)
                    (2, 1.0)    
                    (3, 0.8) 
                    (4, 0.6) 
                    (5, 1.0) 
                    (6, 0.8) 
                    (7, 2.0) 
                    (8, 0.6) 
                };
        \end{axis}
    \end{tikzpicture}
    \end{adjustbox}
    \caption{Visual representation of different pipelines in terms of both RobustQA Average score and Runtime.}
    \label{fig:graph}
\end{figure}

\section{Conclusion}

\autoref{fig:graph} visually represents the comparative analysis of different search and retrieval methods integrated with language models. It juxtaposes the RobustQA average score (in blue bars) and average response time (in red line with markers) for each method. This visualization aids in understanding how each method performs in terms of accuracy and efficiency. The Graph search algorithm with LLM and Retrieval awareness (Writer Retrieval) stands out as the most effective method, balancing high accuracy with quick response times. LlamaIndex with Weaviate Vector Store is also notable for its high accuracy. On the other hand, RAG implementations (both Google Cloud and Amazon SageMaker) lag in performance. This analysis suggests a trend where specialized retrieval-aware methods combined with efficient language models lead to better performance in both accuracy and response time.

%%%%%%%%%%%%%%%%%%%%%%%%%%%%%%%%%%%%%%%%%%%%%%%%%%%%%%%%%%%%

\bibliographystyle{acl_natbib}
\bibliography{palmyra_med}

\end{document}